# Real-Time Impulse Noise Removal from MR Images for Radiosurgery Applications

Zohreh HosseinKhani[1], Mohsen Hajabdollahi[1], Nader Karimi[1], S.M. Reza Soroushmehr[2,3], Shahram Shirani[4], Shadrokh Samavi[1,4], Kayvan Najarian[2,3]

[1]Department of Electrical and Computer Engineering, Isfahan University of Technology, Isfahan 84156-83111, Iran
[2]Michigan Center for Integrative Research in Critical Care, University of Michigan, Ann Arbor, MI 48109 U.S.A
[3]Department of Emergency Medicine, University of Michigan, Ann Arbor, MI 48109 U.S.A
[4]Department of Electrical and Computer Engineering, McMaster University, Hamilton, ON L8S 4L8, Canada

## Abstract

**Background and objective:** In the recent years image processing techniques are used as a tool to improve detection and diagnostic capabilities in the medical applications. Medical applications have been so much affected by these techniques which some of them are embedded in medical instruments such as MRI, CT and other medical devices. Among these techniques, medical image enhancement algorithms play an essential role in removal of the noise which can be produced by medical instruments and during image transfer. It has been proved that impulse noise is a major type of noise, which is produced during medical operations, such as MRI, CT, and angiography, by their image capturing devices. An embeddable hardware module which is able to denoise medical images before and during surgical operations could be very helpful.

**Methods:** In this paper an accurate algorithm is proposed for real-time removal of impulse noise in medical images. All image blocks are divided into three categories of edge, smooth, and disordered areas. A different reconstruction method is applied to each category of blocks for the purpose of noise removal.

**Results:** The proposed method is tested on MR images. Simulation results show acceptable denoising accuracy for various levels of noise. Also an FPAG implementation of our denoising algorithm shows acceptable hardware resource utilization. Hence, the algorithm is suitable for embedding in medical hardware instruments such as radiosurgery devices.

## 1. Introduction

Medical images are affected by different types of noise. Presence of impulse noise can produce misleading artifacts in the visual representation of the interior of human organs. These artifacts could mislead the expert in the process of diagnosis or prognosis. Noise can be produced in different types of medical image instruments such as MR, CT, X-ray, ultrasound, etc. In medical applications the probability of noise creation is increased due to the fast scanning process [1]. Different types of noises in medical imaging instruments, such as impulse, Gaussian, and speckle, can be created during image capture or transmission [1,2,3,4]. Numerous research works have been conducted in detection and elimination of noise in medical images. In [1] an adaptive filter with edge preserving property for Rician noise in MRI images is proposed. In [2], for Gaussian and impulse noise detection in tomography images, a discriminative bilateral filtering for is proposed. In [3] an adaptive median filter for removal of impulse noise in X-ray images and speckle noise in ultrasound images is proposed. In [4] medical images which are used for detecting cancer in different parts of human body are considered and different types of noise effecting such images are reviewed.

Impulse noise is of interest to many researches as a type of common noise. Most denoising methods that are proposed for impulse noise in natural images are computationally complex. For example, fuzzy methods in colored videos, evolutionary algorithms, and an uncertainty based detector with a weighted fuzzy filter, which are used in [5-7], can be considered as denoising methods with high complexity. On the other hand, some denoising methods have lower complexity. For example, in [8-11] a patch oriented approach, based on the image texture, is used for noise detection. In [8, 10] median operation is used for image reconstruction and in [9] a reconstruction method, based on edge directions, is considered. In [11] median operation on non-noisy pixels, is performed as the restoration method. Enhancing the MR images is of importance in segmentation of the gray matter of the brain. With the advancement of the image-guided surgical approaches, segmentation of MR images is becoming an important tool [12]. Therefore MR image enhancement and denoising play essential roles before and during surgical operations such as radiosurgeries. Denoising algorithms have two major steps: a) noise detection, b) noise removal.

Many studies have tried to enhance and remove the impulse noise in MR images. In [13], a wavelet network,

as a preprocessing stage, and a median operation for removal of noisy pixels in medical images are proposed. Differences between gray-scale values and the average of the background are feed to a wavelet network. This is done to detect the location of the noise and the median operation is used for reconstruction of the noisy pixels. In [14] a hybrid filtering method using median operation, based on the structure of each image block, is proposed for impulse noise removal. In [15] a fuzzy genetic algorithm is proposed which has relatively high computation complexity. For denoising filters designed by genetic algorithm and to obtain optimum filters, fuzzy rules are used to modify the cross over probability. Although noise in MR images can be reduced by averaging between multiple imaging results, but it increases image acquisition time. Therefore, in most cases a filtering method as a post processing step is used. The filtering process may involve image blurring and smoothing [16].

There are numerous studies to accelerate medical image processing algorithms using hardware-accelerators, such as FPGAs and GPUs [17]. Since MR image processing can be a time consuming task, thus hardware implementation of these algorithms can be beneficial to obtain a better performance [17].

The need for real-time implementation of some enhancement techniques, such as denoising, makes hardware implementation more appealing. In [18], for the detection step, the maximum and minimum intensities in a 3×3 window are calculated. For the restoration step, edge directions are considered and noisy pixels are restored in the correct edge direction. In [19] a real-time approach for the removal of impulse noise is considered. A noisy pixel detection method and a weighted filtering method are implemented for the reconstruction of noisy pixels. In [20] a decision-tree and similarity between neighboring pixels are proposed. Similar to the algorithm of [18], edge direction is utilized in order to restore the noisy pixels.

In the denoising process, different factors such as accuracy, scalability and complexity must be taken in to account. All of the mentioned factors are important, but in some applications some of these properties become critical. In real-time applications and for embedding an algorithm in a medical imaging instrument, it is essential to decrease complexity and increase the speed of operation.

In this paper we are proposing an accurate and real-time algorithm to detect and remove impulse noises in MR images. For local image blocks different structures, such as edges, smooth, and disordered areas, are considered. Different reconstruction methods are applied for each block depending on the noisy pixels. Due to efficient detection and adaptive reconstruction method, noisy pixels are removed while image structures are preserved accurately. This type of performance is essential in processing of medical images. All steps of the proposed denoising algorithm are designed to have low complexity. For each part of the proposed algorithm a hardware implementation is designed and optimized which makes the proposed method suitable for denoising of images in medical imaging instruments.

The rest of this paper is organized as follows. The proposed method for removal of random value impulse noise, composed of software algorithm and hardware architecture are explained in section 2 and section 3, respectively. Section 4 is dedicated to simulation results, and after that, in section 5 concluding remarks are presented.

## 2. Proposed method

Impulse noise is a common type of noise, which is randomly distributed throughout the image. Impulse noises are divided into two main types, according to the range of the injected values.

A) Fixed value impulse noise (FVIN)

As shown in (1), in grayscale images the randomly injected values could belong to one of the two constant ranges [17]. In (1), $m$ is a constant value, $x_{i,j}$ and $y_{i,j}$ are original and noisy value of the pixel respectively, and $p1$ and $p2$ are the probabilities that the pixel gets the noisy value, where $p = p1 + p2$. Also, $(L-1)$ shows the maximum possible intensity value of a pixel.

$$y_{i,j} = \begin{cases} [0, m) & \text{with probability } p1 \\ x_{i,j} & \text{with probability } (1-p) \\ [(L-1)-m, (L-1)] & \text{with probability } p2 \end{cases} \quad (1)$$

If $m = 0$ then the induced noise is salt and pepper noise.

B) Random value impulse noise (RVIN)

As shown in (2) for gray scale images, a pixel may get noisy with a probability of $p$, where a value in the range of 0 to $(L-1)$ is randomly chosen to replace the original pixel [18]. In (2), $r$ is a random value, $x_{i,j}$ and $y_{i,j}$ are the original and new values of the pixel respectively.

$$y_{i,j} = \begin{cases} r & \text{with probability } p \\ x_{i,j} & \text{with probability } 1-p \end{cases} \quad (2)$$

In this research we are considering RVIN, which is more common noise and its removal is more challenging. For random value impulse noise it is not easy to label a pixel as being noisy because it could have any grayscale value. To overcome this issue in the proposed method we categorize all image blocks in five block types. These categories are called smooth, noisy-smooth, edge, noisy-edge, and disordered blocks.

### 2.1. General structure of the algorithm

The block diagram of the proposed method, a sample noisy image, and the restored image, are displayed in Fig. 1

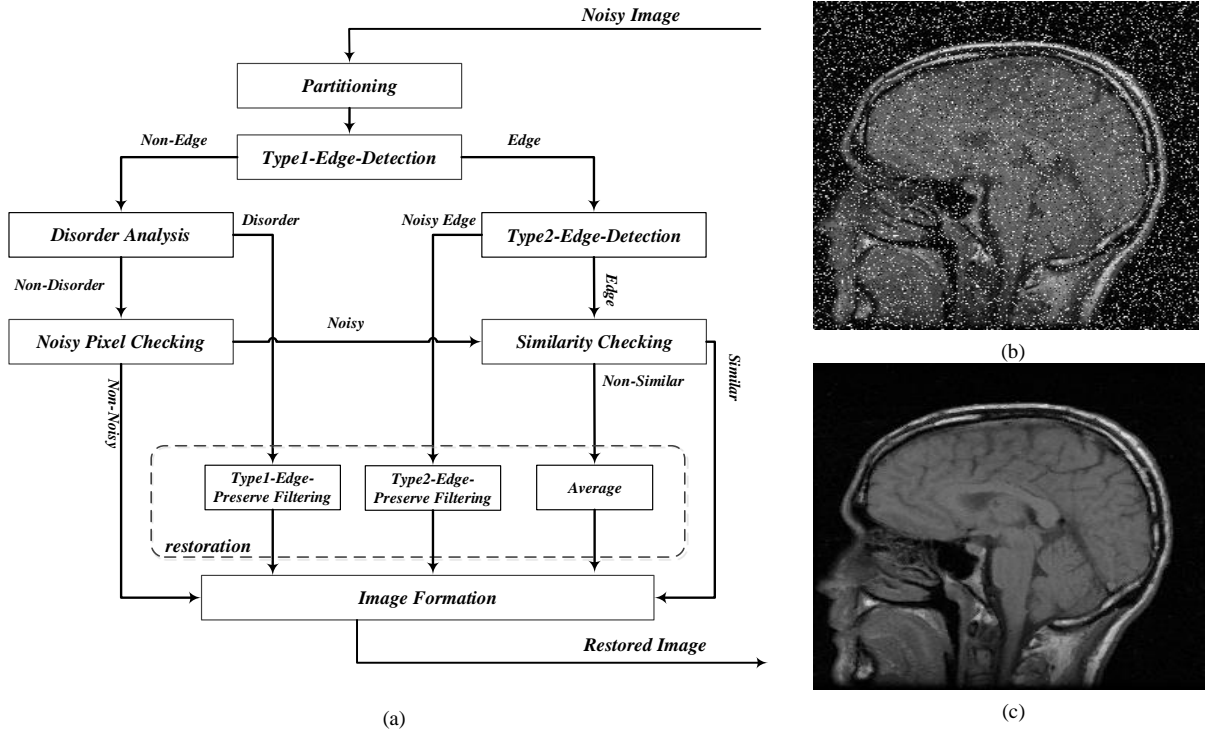

Fig. 1. (a) General structure of the proposed algorithm, (b) a sample noisy image, (c) the denoised image.

As illustrated in Fig. 1, for each detected block, different reconstruction methods are used. For better illustration of our proposed method, the algorithm is applied to four different image blocks and the results are illustrated step by step in Fig. 2. Different stages of the algorithm are as following.

### 2.2. Partitioning

Normally, neighboring pixels of an image block have similarities. Such characteristic diminishes in the presence of noise. To detect abnormal variations in pixel values, the neighborhood of each pixel should be analyzed to find out if the pixel is a normal part of that neighborhood or not. Hence, the proposed method partitions the image into 3×3 and 5×5 blocks depending on the local structure of the image and the severity of the noise. It is necessary to use variable block size for a better detection in different image blocks. For example, in edge blocks, the larger block size leads to better edge detection and better image restoration. In this paper the 9 pixels of the 3×3 block are called $P_1, P_2, ..., P_9$, where $P_5$ is the central pixel. Also in 5×5 blocks the pixels are called as $P_1, P_2, ..., P_{25}$, where $P_{13}$ is the central pixel. As illustrated in row 1 of Fig. 2, in order to find a better detection of block categories, pixel intensities are sorted. In this paper the sorted pixel values are called $F_1, F_2, ..., F_9$. Partitioning of pixels in different size blocks can be useful for the next processing stages which are discussed in the following subsections.

### 2.3. Edge Detection

In noisy conditions, edges blocks can be affected by noise, hence it is necessary to find out they are noisy or non-noisy. Here edge detection is done in two steps including *Type1-Edge-detection* and *Type2-Edge-detection*. In the first step using *Type1-Edge-detection* it is determined that a block is containing an edge or not. This step is checked at first in all image blocks as illustrated in row 2 of Fig.2. In the second step using a 5×5 block with respect to the edge directions, rough and non-noisy edges are separated from noisy edges. These two steps of edge detection are explained as follows:

#### 2.3.1. Type1-Edge-Detection

A 3×3 block around each pixel is considered and elements of the block are sorted. The differences between the 5th and 4th sorted elements, $(F_5 - F_4)$, or the 5th and 6th elements, $(F_6 - F_5)$, represent the edge strength in the block. Then using a threshold $(T_1)$ the central pixel is labeled as edge based in Equation (3).

$$P_5 = \begin{cases} edge \ if \ (F_5 - F_4) \ or \ (F_6 - F_5) > T_1 \\ non-edge \ otherwise \end{cases} \quad (3)$$

As illustrated in row 2 of Fig. 2, *Type1-Edge-Detection* is performed as first stage on all image blocks. In each step in row 2 of Fig. 2, green and red colors indicate that the criterion for this step is met or not, respectively. For blocks that the condition is true *Type2-Edge-Detection* is also performed and for blocks that *Type1-Edge-Detection* is false Disorder-Analysis is performed. *Type2-Edge-Detection* and *Disorder-Analysis* which are considered as step 3 of the proposed algorithm are discussed in the following.

### 2.3.2. Type2-Edge-Detection

For all pixels that are labeled as edge by the *Type1-edge-detection*, the second edge detection criterion is also checked. In *Type1-Edge-Detection* all edges are detected but it is not known whether these edges are noisy or not. In this step, noisy pixels are detected by considering major edge directions in a 5×5 block. To do so, as shown in Fig. 3, four main directions of horizontal, vertical, diagonal, and anti-diagonal, are considered. In each of the four directions the weighted sum of absolute differences, $D_i|_{i=1,\dots,4}$, between the central pixel and the other pixels located on a particular direction, is calculated based on Equation (4).

$$D_i|_{i=1,\dots,4} = \sum_{j=-2,-1,1,2} |I_c - W_j I_j| \quad (4)$$

where $I_c$ is the central pixel, $I_{\pm 1}$ are the two pixels that are closest to the central pixel in each direction. Also, $I_{\pm 2}$ are the two pixels that are farthest from the central pixel, in each direction. For the two farthest pixels of $I_{\pm 2}$ a weight coefficient of ½ is considered, which means $W_j = \frac{1}{2}|_{j=\pm 2}$. For the two nearest pixels of $I_{\pm 1}$ a weight coefficient of 1 is considered, which means $W_j = 1|_{j=\pm 1}$. This operation is done in all four main directions. The minimum value, in all four directions, $D_{min} = min(D_i|_{i=1,\dots,4})$, shows the most probable edge direction. If $D_{min}$ were to be less than a threshold $(T_2)$ there is a high similarity between the central and the edge pixels, and the central pixel is considered as an edge pixel. However, if $D_{min} > T_2$, it would be labeled as a noisy edge pixel. In Fig. 2, two examples of edge and noisy edge blocks are shown in columns 1 and 3 of row 3, respectively. According to *Type2-Edge-Detection*, in noisy edge blocks (in column 3 of row 3) $D_{min}$ is greater than $T_2$, where $T_2$ is considered equal to 150. This situation of noisy edge block is shown by a red colored block. Noisy edge blocks are labeled to be fed into *Type2-Edge-Preserve Filtering* step, which is discussed later. On the other hand, non-noisy edge occurs (in column 1 of row 3) if $D_{min} \leq T_2$. This situation is colored by green. Non-noisy edge blocks are labeled and are fed into the *Similarity-Checking* step which is discussed section 2.6.

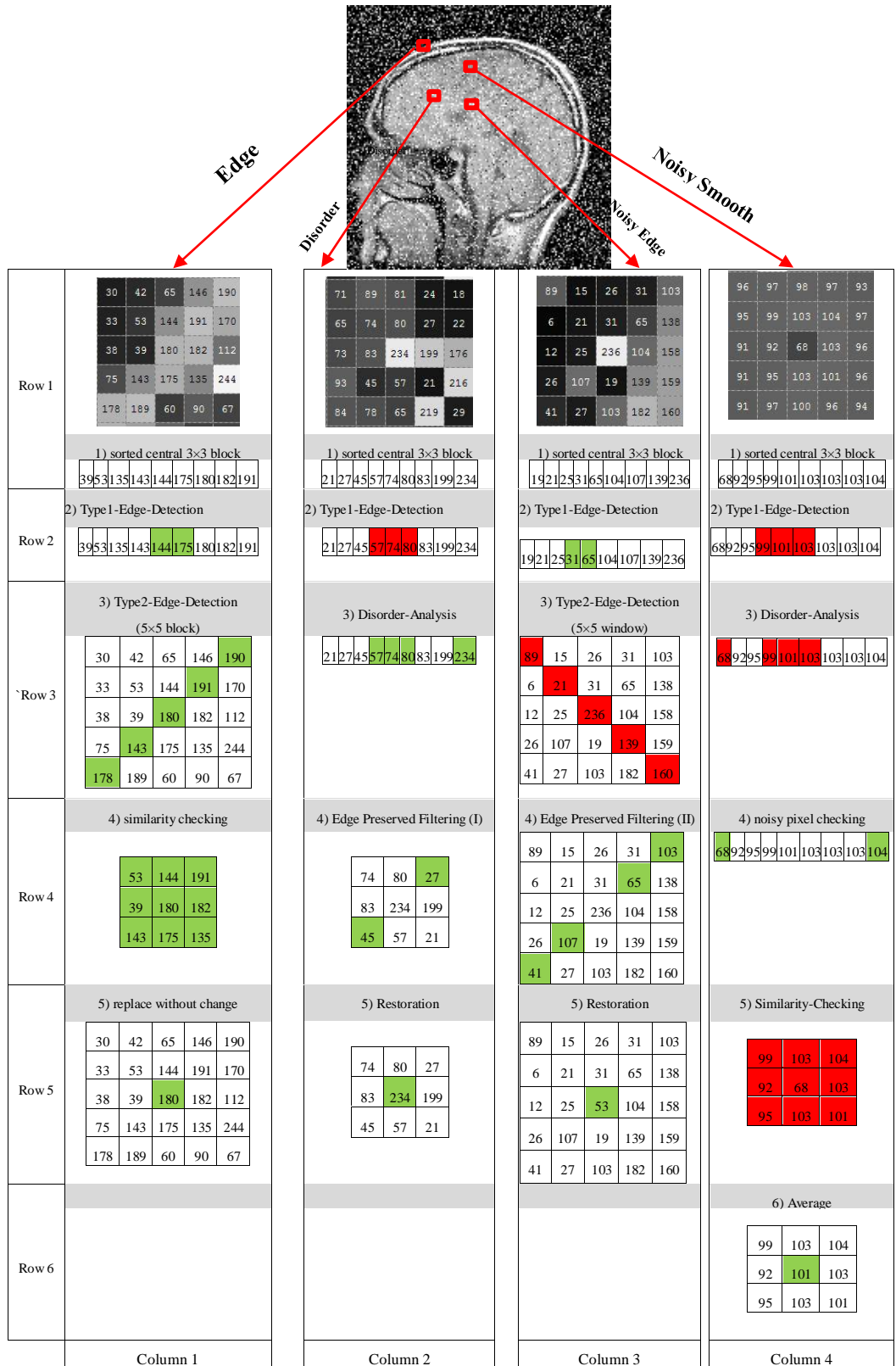

Fig. 2. Example of noise detection and image reconstruction procedure for different image localities.

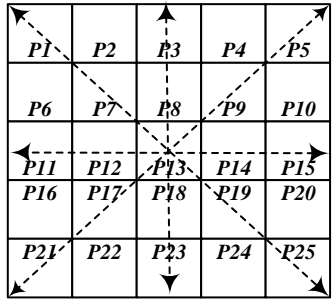

Fig. 3. Pixel numbers and edge directions in *Type2-Edge-Detection* window.

## 2.4. Disorder Analysis

When a pixel is labeled as non-edge by *Type1-Edge-Detection*, it is important to know whether the pixel is in a smooth or in a disordered area. In the *Disorder-Analysis* step those non-smooth blocks, and blocks that their central pixels have different values from their surrounding pixels, are considered as disordered blocks. Considering the sorted pixels of the $3 \times 3$ *Type1-Edge-Detection* window we need to work with the three median ones. Absolute difference between the central pixel and three median sorted pixels including $F_4$, $F_5$ and $F_6$ is a measure of disorder within the $3 \times 3$ neighborhood.

$$P_5 = \begin{cases} Disordered & if \ |F_6 - P_5| \ and \\ & |P_5 - F_4| \ and \\ & |P_5 - F_5| > T_3 \\ smooth & otherwise \end{cases} \quad (5)$$

where $T_3$ is a threshold value. Details of the disorder analysis procedure are visually represented in columns 2 and 4 of row 3 of Fig. 2. The 4th, 5th, and 6th sorted elements are considered in Fig. 2, and their differences with the central pixel is compared with threshold $T_3$. If all three absolute differences are greater than $T_3$, the image block is considered to be disordered. Since the disordered blocks are potentially expected to be noisy, the *Noisy-Pixel-Checking* procedure is applied as the next step. As illustrated in Fig. 2 in columns 2 and 4 of row 3, for disordered blocks, the central pixel as well as three sorted pixels ($F_4$, $F_5$, $F_6$) are highlighted with green. If the condition of Equation (5) is not met then the mentioned pixels are highlighted as red.

## 2.5. Noisy Pixel Checking

In the disorder analysis, blocks which were detected as smooth, may be contained the noisy pixels. In fact these blocks are included noisy pixel surrounded by smooth pixels. Hence, in a smooth area, those pixels which have different intensity values from their background are determined as noisy pixels. As Equation 6 the differences between the central pixel $P_5$ and the maximum or minimum pixels inside the 3×3 window are used for detection of a noisy pixel.

$$P_5 = \begin{cases} noisy & if \ (F_9 - P_5) \ or \ (P_5 - F_1) < T_4 \\ not \ noisy & otherwise \end{cases} \quad (6)$$

If either $(F_9 - P_5)$ or $(P_5 - F_1)$ is less than a threshold $T_4$ then the pixel is considered as noisy pixel in a smooth area. In some conditions all non-noisy block pixels may have nearly maximum or minimum values. In such a case they are wrongly considered as noisy pixels based on Equation 6. To prevent this wrong decision, similarity between a noisy pixel and its neighbors is checked by the *Similarity-Checking* step.

## 2.6. Similarity Checking

Checking out the block similarity is necessary in two conditions. First when we want to leave the pixel without any modification. In row 4 of column 1 in Fig. 2 non-noisy edges are replaced without any modification. Second when by only using the pixel's intensity the noisy pixel is detected as illustrated in row 5 of column 4 in Fig. 2. Hence, in Fig. 2, similarity must be checked when *Type2-Edge-Detection* and *Noisy-Pixel-Checking* have true conditions. Absolute differences between the central pixel and its eight neighbors are calculated to determine the similarity amount. Using threshold $T_4$, these absolute differences determine similarity or non-similarity among these 8 pairs. If the number of the similar pixel around a pixel becomes less than a threshold ($T_5$), then it is considered to be a noisy pixel. In Fig. 2, if central pixel is similar with its 3×3 neighbors, thus all pixel blocks are colored with green (row 4 of column 1) otherwise they are colored with red (row 5 of column 4).

## 2.7. Restoration

The restoration mechanisms are different in different blocks. Three methods for restoring the original pixel value are proposed including averaging on fourth, fifth and sixth elements of sorted results (the *Average* method), *Type1-Edge-Preserve Filtering* and *Type2-Edge-Preserve Filtering*. Type of the restoration method depends on the block in which the pixel is detected in. Three restoration methods are as follows.

### 2.7.1. Average

In smooth blocks as well as in blocks that a pixel has similar value to its neighbors, restoration is performed by averaging on fourth, fifth and sixth elements of sorted 3×3 block as depicted in Fig. 2 (row 6 of column 4).

### 2.7.2. Type1-Edge-Preserve Filtering

In this step the noisy pixels are restored using the direction of the edge. In [20] details of applied *Type1-Edge-Preserve Filtering* are presented. As illustrated in Fig. 2 (row 5, column 2), in *Type1-Edge-Preserve Filtering,* for disorder type blocks, two pixels that are used in the averaging process are colored with green.

### 2.7.3. Type2-Edge-Preserve Filtering

Noisy pixels which are detected in edge areas are restored with *Type2-Edge-Preserve Filtering*. In a 5×5 block, four main directions including horizontal, vertical, diagonal, and anti-diagonal, are considered. All pixels corresponding to each direction are considered. Sum of absolute differences between each pixel and their corresponding average is calculated. In this step central pixel isn't considered in final results because this pixel is a noisy pixel. Next to determine the possible direction of the edge, the minimum value in four directions is computed. Finally restoring is performed by taking a median operation on directions which is determined in the previous step. As illustrated in Fig. 2 (row 4 of column 3), pixels on possible direction of the edge used for median restoration are colored with green.

### 2.8. Image Formation

Noise-free pixels detected in the previous steps as well as restored pixels are placed back in order to form the noise-free image.

## 3. Hardware architecture

The proposed noise removal algorithm is designed to be suitable for hardware implementation. As illustrated in Fig. 4, a 3×3 window around each pixel is considered and its elements are sorted by a sorting module. Results of the sorting module are feed to four computational modules, including *Disorder-Analyzer*, *Type1-Edge-Detection*, *Noisy-Pixel-Checker* and a module which performs averaging of the three medians of sorted elements (the *Average* method).

Different parts of the hardware structure of the proposed algorithm are explained in the followings:

### 3.1. Edge Detection Module (I)

Two subtractors, two comparator units, and a logical OR gate form the circuit which can be used as the *Type1-Edge-Detection* module as shown in Fig. 5. It could be used for implementation of (6) as a *Noisy-Pixel-Checking* module by changing the inputs of the circuit.

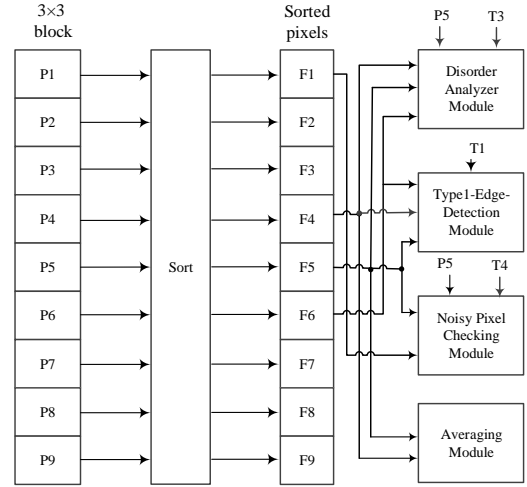

Fig. 4. Sorter module and modules that feed from it.

### 3.2. Type2-Edge-Detection Module

In Fig. 6, the hardware structure of this module is illustrated. In *Type2-Edge-Detection*, absolute differences of the central pixel with corresponding pixels located on edge direction are computed and weighted sum of them is computed. Sixteen absolute difference calculation modules (ABS-DIF) as well as eight shift register which make weighted results are used. After that an adder and a comparator is enough to detect the edge direction in this module.

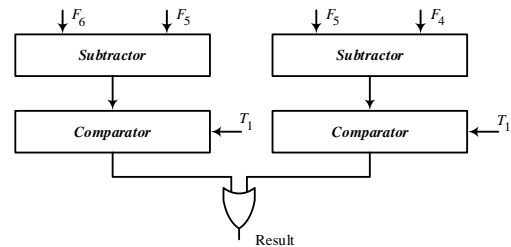

Fig. 5. Hardware realization of *Type1-Edge-Detection* module.

### 3.3. Similarity-Checker Module

In Fig. 7 we are using eight absolute-difference-calculation modules (ABS-DIF). They are used to calculate the absolute differences between the central pixel and its 8 neighbors. Afterward the results are compared with the threshold $T_4$, by using eight comparator units. A positive result from each comparator indicates the similarity of that pixel with the central pixel. Finally sum of similar pixels are added by an adder unit. The output of the adder is compared with threshold $T_5$ to produce the similarity output.

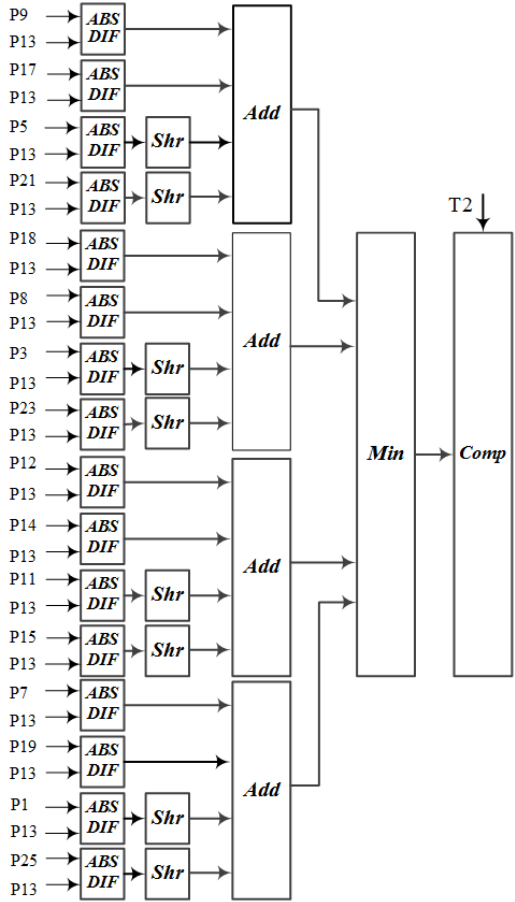

Fig. 6. *Type2-Edge-Detection* module.

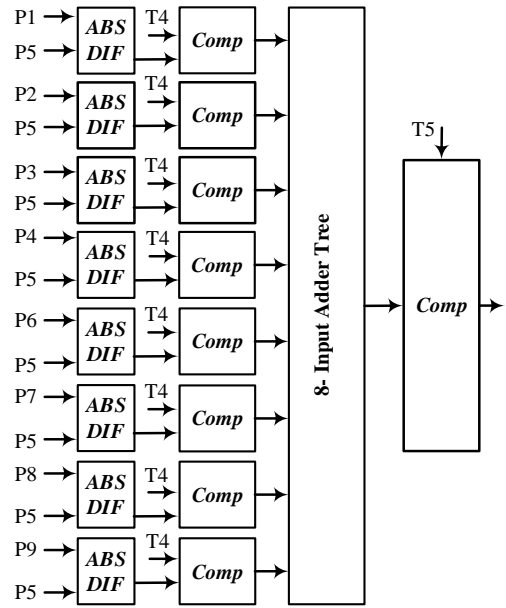

Fig. 7. *Similarity-Checker* module.

## 3.4. Disorder-Analysis Module

In Fig. 8 disorder-analysis-module is shown. Three absolute-difference-calculation modules (ABS-DIF) are used for calculation of the absolute difference between central pixel and three medians of the sorted elements. Then these values are compared with threshold ($T_3$) using comparator module. Final result is provided by logical AND operation of the comparator outputs.

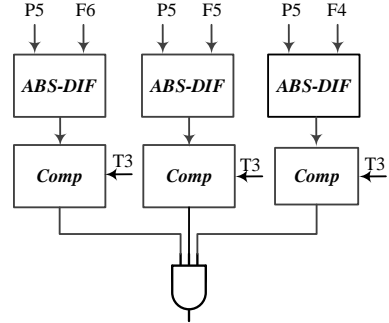

Fig. 8. Hardware realization of *Disorder-Analysis* module.

## 3.5. Edge Preserved Filtering (II)

In the variance-calculation-module (VAR) four ABS-DIF units are used for each main direction. These four units are for calculating the absolute difference between each edge pixel and the average value of the pixels in that direction. These four differences are added by an adder module. Figure 9 shows one VAR unit for the anti-diagonal direction. At the next step, as illustrated in Fig. 10, the minimum of the four variances selects one of the four inputs of a multiplexer. Multiplexer inputs are the medians of the four directions. Hence, the direction with least variance is selected and the median of that direction replaces the noisy pixel.

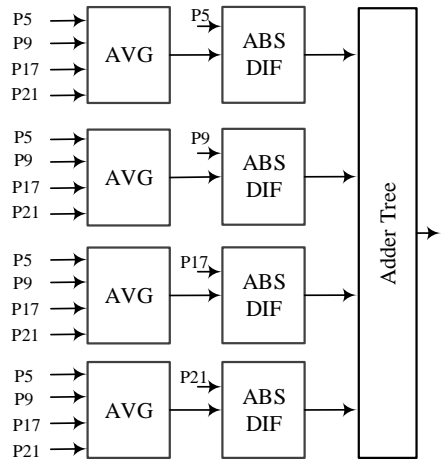

Fig. 9. Hardware realization for one of the four VAR modules.

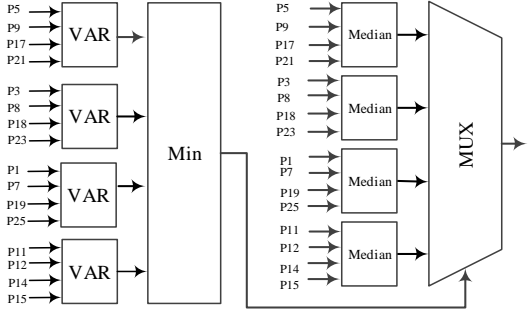

Fig. 10. Hardware realization of *Type2-Edge-Preserve Filtering* module.

### 3.6. Other hardware module

For sorting module, a simple structure of [22] is employed. Also, for the *Type1-Edge-Preserve Filtering*, the hardware architecture of [20] is applied. In the image formation step, the restored and non-noisy pixels are replaced in proper locations based on the type of pixel. Pixel value replacement can be performed by a multiplexer or by simple wiring.

## 4. Experimental Results

The proposed method's accuracy and complexity are verified by simulation in the following two stages:

### 4.1. Software Simulation

Experiments are performed and verified with MATLAB. In this study, 124 standard 8-bit gray-scale MR images with the size of 256×256 are used [23]. Noise densities between 5% and 40% are uniformly injected. Objective testing of peak signal-to-noise ratio (PSNR) is used to assess the quality of the restored images. In our proposed method we set thresholds $T_1 =$ 20, $T_2=150$, $T_3=30$, $T_4=10$ and $T_5=6$ and in order to achieve better results, the algorithm was repeated twice. In the first iteration due to high noise levels, no similarity can be observed by the *Similarity-Checking* stage. Hence, the *Noisy-Pixel-Checking* module does not function. Two hardware architectures, proposed in references [8, 20], are used for removal of the impulse noises. Also, $3 \times 3$ and $5 \times 5$ median filters are used for comparison. As shown in the Table I, the proposed method has better results than the compared methods for all noise densities.

To show some visual results of the proposed method, in Fig. 11 four original MR images and their noisy versions with the presence of 20% impulse noise are shown. In Fig. 12 median filter is used for noise removal and PSNR (dB) values are reported for each image. In Fig. 13 comparison of the proposed method with [8] and [20] are shown. Simulation results, as shown in Fig. 13, indicate that the proposed method produces better image qualities in terms of PSNR values.

### 4.2. Complexity Analysis

For complexity analysis of the proposed method, FPGA implementation is investigated. The proposed architecture is described in VHDL and is implemented on a XILINX virtex4 family xc4vfx12 device. Implementation specifications as well as average PSNR, for noise densities of 5%, 10%, 15% and 20%, are reported and compared with the other studies in Table II. It clearly can be seen that the proposed method has a better image quality and it has acceptable hardware implementation results.

**Table I** Comparison between denoised results in terms of PSNR (dB) for different noise densities.

|  | 5% | 10% | 15% | 20% | 30% | 40% |
|---|---|---|---|---|---|---|
| **Median3×3** | 34.27 | 33.17 | 31.14 | 28.40 | 22.89 | 18.41 |
| **Median 5×5** | 30.18 | 29.97 | 29.66 | 29.28 | 27.76 | 23.77 |
| **LCNR[8]** | 38.27 | 35.65 | 32.18 | 28.65 | 22.67 | 18.13 |
| **Ref [20]** | 36.18 | 34.93 | 33.78 | 32.48 | 29.62 | 26.18 |
| **Proposed Method** | 38.11 | **36.61** | **35.27** | **33.96** | **30.90** | **26.44** |

**Table II** Comparison on implementation specifications between proposed method and methods of [8] and [20].

| Method | Target Device | Area | Delay (ns) | Average PSNR in 5%, 10%, 15% and 20 % noise |
|---|---|---|---|---|
| **LCNR [8]** | Altera Cyclone II EP2C20F484C7N | 513 (Logic cell) | 7.72 | 33.68 dB |
| **Ref [20]** | Altera FLEX10KE EPF10K200-SRC240-1 | 2166 (Logic cell) | 14.90 | 34.34 dB |
| **Proposed Method** | Xilinx virtex4 xc4vfx12 -12-sf363 | 2761 (Slice) | 49.25 | 35.98 dB |

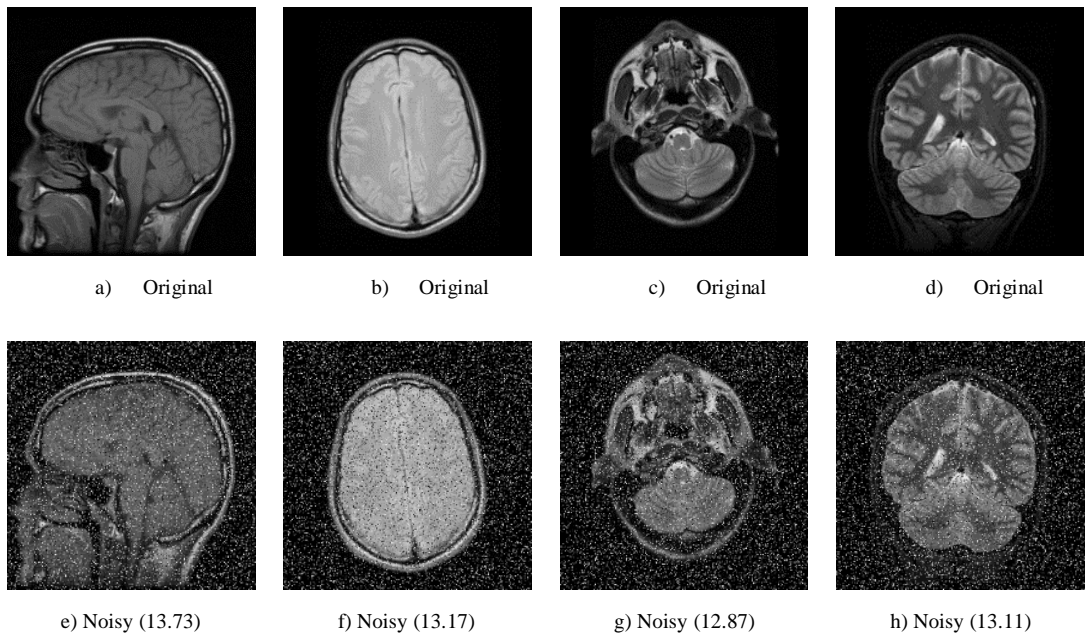

a) Original  b) Original  c) Original  d) Original

e) Noisy (13.73)  f) Noisy (13.17)  g) Noisy (12.87)  h) Noisy (13.11)

Fig. 11. Four sample original images and their noisy versions. PSNR (dB) values mentioned for noisy images.

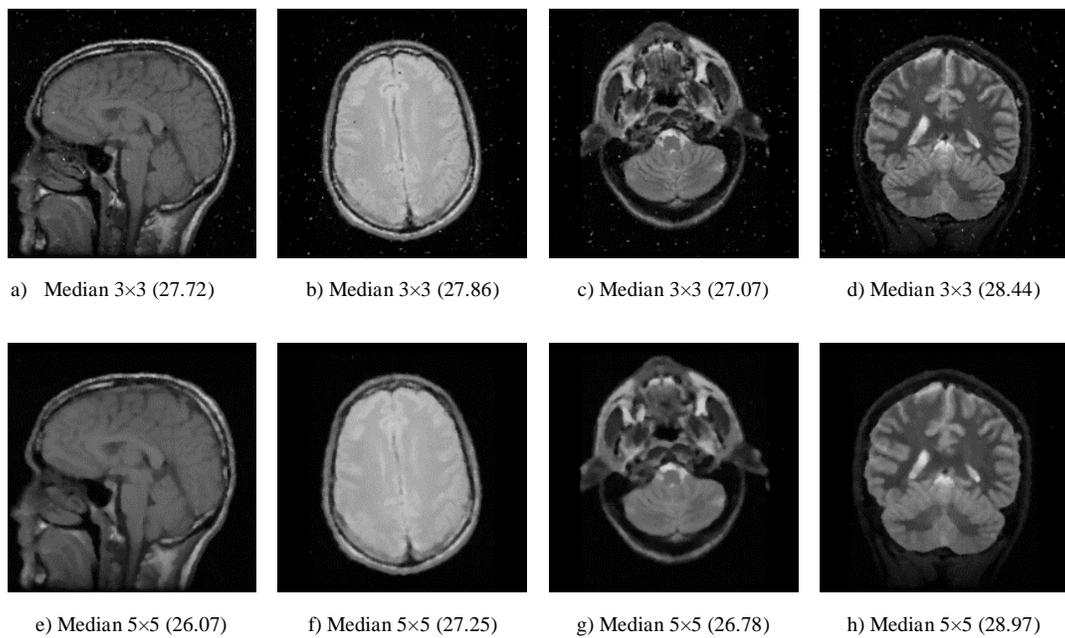

a) Median 3×3 (27.72)  b) Median 3×3 (27.86)  c) Median 3×3 (27.07)  d) Median 3×3 (28.44)

e) Median 5×5 (26.07)  f) Median 5×5 (27.25)  g) Median 5×5 (26.78)  h) Median 5×5 (28.97)

Fig. 12. Denoising of images of Fig. 11 using standard median filters. PSNR (dB) values are indicated for denoised images.

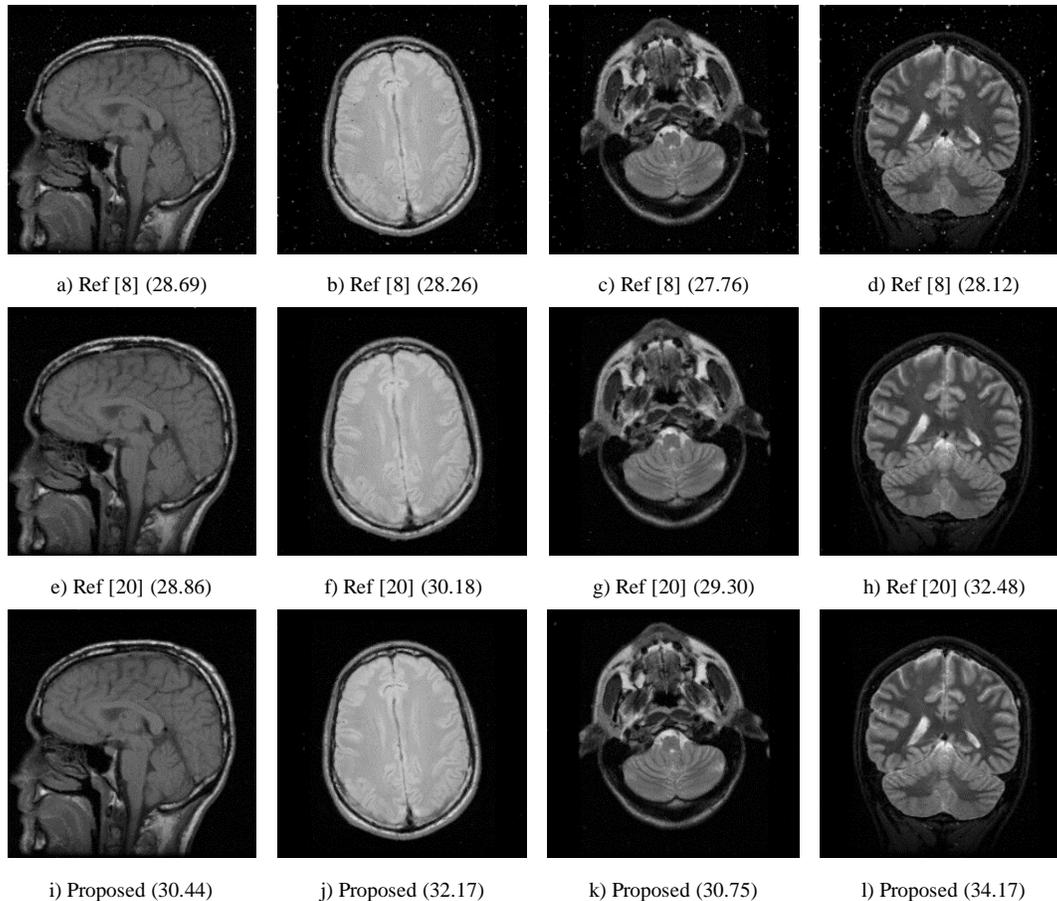

a) Ref [8] (28.69)  b) Ref [8] (28.26)  c) Ref [8] (27.76)  d) Ref [8] (28.12)

e) Ref [20] (28.86)  f) Ref [20] (30.18)  g) Ref [20] (29.30)  h) Ref [20] (32.48)

i) Proposed (30.44)  j) Proposed (32.17)  k) Proposed (30.75)  l) Proposed (34.17)

Fig. 13. Visual quality comparison of the proposed method with [8] and [20]. PSNR (dB) values are also shown.

## 5. Conclusion

In this paper a low complexity noise removal system for MR images was implemented. This method was proved to be suitable for hardware implementation. Such implementation can be used as a part of the medical image capturing instruments for enhancement and segmentation of MR images before and during of surgical operation in radiography and radiosurgery. The mentioned method contained two steps of detection and restoration. The achieved goal was the enhancement of the accuracy in each stage separately. High accuracy of noisy-pixel detection in the first stage, and their removal in the next stage, led to a better restoration of noisy images. Simulation results using MATLAB software, performed on MR images, demonstrated that the proposed approach removed random value impulse noise with high accuracy. Also, FPGA implementation of the proposed method resulted in low hardware resource utilization and produced high quality denoised images.